\documentclass[aps,twocolumn,groupedaddress]{revtex4}

\usepackage{epsfig,amsbsy,bm}

\newcommand{\eref}[1] {(\ref{#1})}

\newcommand{\etal}{{\em et al.  \,}}
\newcommand{\isum}%
{\mathop{\hbox{$\displaystyle\sum\kern-13.2pt\int\kern1.5pt$}}}

\renewcommand{\k}{{\bm k}}
  \newcommand{\q}{{\bm q}}
\renewcommand{\r}{{\bm r}}

\begin{document}
\bibliographystyle{apsrev}


\title
{Convergent calculations of  double ionization of helium:\\ from
($\gamma$,2e) to  (e,3e) processes}

\author{A. S. Kheifets}
\affiliation{Research School of Physical Sciences,
The Australian National University,
Canberra ACT 0200, Australia}

\author{Igor Bray}
\affiliation{Centre for Atomic, Molecular, and Surface Physics,
         School of Mathematical and Physical Sciences,
         Murdoch University, Perth, 6150 Australia
}

\date{\today}

\begin{abstract}
The first absolute (e,3e) measurements, by Lahmam-Bennani \etal
[Phys.~Rev.~A {\bf 59}, 3548 (1999)], have been recently approximately
reproduced by Berakdar [Phys.~Rev.~Lett.~{\bf 85}, 4036 (2000)] and
supported by Jones and Madison [Phys.~Rev.~Lett.~{\bf 91}, 07321
(2003)], but with widely differing conclusions. The former indirectly implied that
the Born-CCC-based calculations of Kheifets \etal [J.~Phys.~B {\bf
32}, 5047 (1999) were invalid due to the reliance on the 1st Born
approximation. The latter argued that the 1st Born approximation was
valid, but the wrong initial state was used. We investigate these
claims and find that the original calculations of Kheifets \etal are
reproduced whether the 2nd Born approximation is incorporated or if we
use a ground state similar to that of Jones and Madison, but
appropriately corrected  as done by Le~Sech and co-workers 
[J.~Phys.~B {\bf 23}, L739 (1990)].
\end{abstract}


\maketitle

When attempting calculations of complex collision processes we have
found it important to utilise a formalism where convergence of the
results can be tested by substantial variation of as many of the input
parameters as possible. The goal is to develop a predictive theory
whose outcomes can be relied upon on scientific merit irrespective of
agreement with experiment or other theory. The convergent
close-coupling (CCC) method \cite{BS92} was developed with this idea
in mind. The total wavefunction is expanded utilising a complete
Laguerre basis and so greater accuracy is ensured with increasing
basis size. In the Coulomb three-body problems, such as e-H
scattering, once convergence is reached there is no freedom left to
alter the results. Hence, when there is disagreement with experiment
as discussed in Ref.~\cite{BS92}, that may be even explained by others
\cite{MT95} or improved upon \cite{D97}, we find that new experiments
\cite{YCC97,Oetal98} are supportive of the original CCC results. The
predictive power of the CCC approach to double photoionization (DPI or
$\gamma$,2e) was demonstrated more recently when initial disagreement
with experiment \cite{KB98l} was resolved in favor of the CCC
calculations
\cite{A01}. 

Whereas in e-H calculations convergence considerations relate
primarily to the usage of the Laguerre basis, in ($\gamma$,2e)
calculations there is additional consideration of convergence with
respect to the description of the initial state. In the more
complicated Coulomb four-body problem that is electron-impact double
ionization of helium (e,3e) there are further considerations of
convergence with respect to the Born approximation order used to treat
the case of a fast projectile. It is these issues that are of interest
to us in the present Letter.

The physics of the He (e,3e) reaction in the very fast projectile mode
where the incident electron has an order of magnitude or more energy
than the two ``slow'' ejected electrons is quite similar to the case
of ($\gamma$,2e). In both cases the initial and final atomic states
are three-body problems of the He$^{2+}$ nucleus interacting with the
two slow electrons. The application of the CCC method to (e,3e)
process under these kinematical conditions is straightforward and
expected to produce results as accurate as those for ($\gamma$,2e).
However, the first Born CCC calculation of (e,3e) on He at 5.6~keV
incident energy \cite{KBDLT99} was found significantly lower in
magnitude (by factors of $\sim3$ and $\sim12$ for 10 and 4~eV ejected
electrons, respectively) as compared with absolute measurements of
\citet{ALB99}. Such a strong disagreement could only be attributed to
deviation from the first Born regime since the treatment of the
initial and final states was the same in the CCC ($\gamma$,2e) and
(e,3e) calculations. Some indications of that followed from the work
of \citet{B00} who reported a good agreement with absolute
measurements \cite{ALB99} on the basis of a lowest-order
implementation of a Faddeev-type approach. These calculations differed
significantly for electron or positron impact (by about a factor of 2)
thus invalidating the first Born approximation.

This conclusion, however, was challenged by the recent work of
\citet{JM03} who managed to get good agreement, both
in shape and magnitude, with the experiment of \citet{ALB99} and the
calculations of  \citet{B00} while staying
entirely within the first Born formalism. As in the first Born
calculations presented in \citet{ALB99}, \citet{JM03} used the
asymptotically exact 3C final state, but with a different
helium ground state. They argued that 
Hylleraas-based ground states, as employed earlier \cite{KBDLT99,ALB99},
fail to satisfy the
Kato cusp conditions \cite{Kato57} and are
inaccurate when the two electrons are close together. In support of
their argument, \citet{JM03} employed the Pluvinage ground state
\cite{Pluvinage50}  which treated the interelectron interaction to all
orders of perturbation theory and satisfied Kato's cusp conditions
exactly. This combination of the 3C final state and the Pluvinage
ground state restored agreement with experiment of \citet{ALB99}
within the first Born model.

In this Letter, we review our earlier (e,3e) calculations
\cite{KBDLT99} in the light of these new theoretical findings
\cite{B00,JM03}. Following the implementation of the 2nd Born term
\cite{K04} we are able to test the contribution of this term for the
kinematics considered here. We have also implemented the Pluvinage
ground state and its improved version as prescribed by
Le~Sech and co-workers \cite{MDleS90,SMleS93}, who argued that the
former lacked the screening of two-particle interactions by the third
particle. Finally, we used a
much bigger CCC expansion, than that used previously \cite{KBDLT99}, in
describing the final state. The new initial and final states are
first applied to helium DPI to check the gauge-dependence as a function
of energy. Subsequently, we consider the
corresponding ($\gamma$,2e) and  (e,3e) cases with two
10~eV outgoing electrons. Given the
relatively low momentum transfer in the (e,3e) experiment we expect it
to be close to the optical limit.

The probability of the (e,3e) reaction is given by the
fully-differential cross-section (FDCS):
\begin{equation}
\label{FDCS}
{d\sigma\over d\Omega^\prime d\Omega_1 d\Omega_2 \, dE_2}=
(2\pi)^4 
\frac{k^\prime k_1k_2}{k_0}
\left|T_{fi}\right|^2,
\end{equation}
where indices 0,1 and 2 are asigned to the incident and two ejected
electrons, respectively.  We treat the projectile as a plane wave and
write the double ionization amplitude
\begin{equation}
\label{born}
T_{fi}=
\frac{4\pi}{q^2}\frac{1}{(2\pi)^3}
\Big< \Psi_{\k_1\k_2}
\Big|
e^{i\q\cdot\r_1}+e^{i\q\cdot\r_2} 
 -  Z\Big|
\Psi_0\Big> 
\nonumber
\end{equation}
as the Fourier transform of the Coulomb interaction
$
V = |\r_0-\r_1|^{-1}+
    |\r_0-\r_2|^{-1} - Z/\r_0 \ 
$
between the projectile and the target. Here $Z=2$ is the nucleus
charge and $\q=\k_0-\k^\prime$ is the momentum transfer. In writing
Eqs.~(\ref{FDCS}-\ref{born}) we use a continuum wave normalization
$\langle \k|\k'\rangle=\delta(\k-\k')$ which gives the incident flux
$j=k_0/(2\pi)^3$.  By setting $\q\rightarrow0$ we reach the optical
limit. In this limit angular correlations of the two ejected electrons
are the same in (e,3e) and ($\gamma$,2e) reactions if we align the
vector $\q$ with the polarization axis of light. The magnitudes of the
corresponding cross-sections differ by a universal scaling factor.

In contrast to the projectile-target interaction, which is treated
here in the first and second orders of the perturbation theory,
the interaction of the two ejected electrons is taken
into account fully.  To obtain the two-electron wave function
$\Psi_{\k_1\k_2}(\r_1,\r_2)$ we employ the convergent close-coupling
(CCC) formalism \cite{KBDLT99}. In brief, the final state wave
function is expanded over the channel functions each of which is a
product of a true continuum state and a positive energy
pseudostate. The latter is obtained by diagonalizing the target
Hamiltonian in the Laguerre square integrable basis.  The multiple
interaction between the two ejected electrons is included by solving
the integral Lippmann-Schwinger equation. Following the successful
application of the CCC formalism to (e,2e) equal energy-sharing cross
sections \cite{B02l} we take Laguerre basis exponential fall-off
parameters to be the same for all $l\le4$ and basis sizes $N_l=60-l$. 
All open plus lowest three closed states were included in the calculations. 
Previously \cite{KBDLT99}, with smaller computational resources, we
were restricted to $N_l=17-l$ and had to vary the exponential fall-off
parameters for each $l\le4$ to avoid inaccuracy associated with
interpolation of the complex amplitudes. Some minor variation of the
present results from those published previously is due to the very
different choices of the Laguerre bases used here.

In our earlier paper \cite{KBDLT99} we 
calculated only the first Born amplitude \eref{born} using two accurate
ground state wave functions: a 15-term multi-configuration
Hartree-Fock and a 20-term Hylleraas.  These two wave functions
produced nearly identical results in the two gauges of the Born operator -
length and velocity. In the present investigation we begin with the
Pluvinage ground state:
\begin{eqnarray}
\label{JM}
\nonumber
\Psi_0(\r_1,\r_2)&=& 
8\pi^{-1}e^{-Z(r_1+r_2)} \
N(k)\ 
\Phi_0(\eta,kr_{12}),
\\\mbox{where}\hspace{1.8cm}&&\\
\nonumber
\Phi_0(\eta,kr_{12}) &=& e^{-ikr_{12}} \ 
_1F_1(1-i\eta,\  2, \ 2ikr_{12}).
\end{eqnarray}
The non-correlated part of the ground state \eref{JM} is the product
of the two hydrogenic orbitals in the field $Z=2$. The correlation
factor $\Phi_0(\eta,kr_{12})$ with $\eta=(2k)^{-1}$ describes the
center-of-mass motion of the electron pair unimpeded by the nucleus.

Le~Sech and co-workers \cite{MDleS90,SMleS93} suggested that the
original Pluvinage ground state \cite{Pluvinage50} could be improved
by introducing the screening of the inter-electron interaction by the
nucleus. This can be implemented by changing the effective strength of
the Coulomb interaction to $\eta=q \ (2k)^{-1}$ where $q<1$. In
addition, the screening of the electron-nucleus interaction by another
electron can be accommodated by introducing the shielding factor
$\cosh(\lambda r_1) + \cosh(\lambda r_2)$ in the non-correlated part
of the ground state with $\lambda<1$.  This shielding effect is well known and can be
incorporated into the simplest non-correlated ground state by
introducing an effective charge $Z-5/16$ where $Z$ is the charge of
the nucleus \cite{BS77}.

\begin{figure}

\epsffile{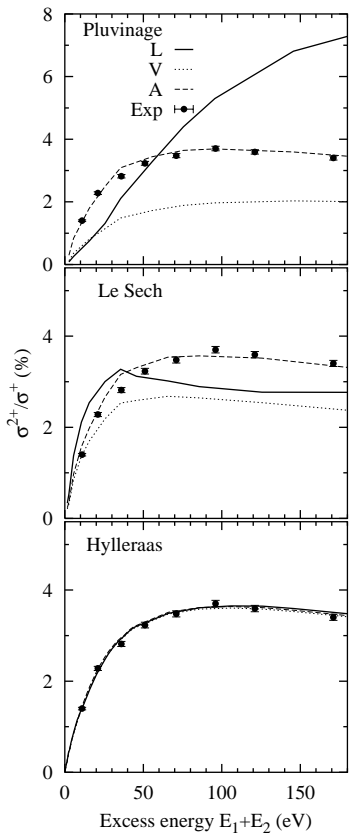}
\caption{
\label{F1}
Double-to-single photoionization cross-section ratio
$\sigma^{2+}/\sigma^+$ as a function of the electron pair energy
$E_1+E_2$. The CCC calculations with different ground states (top -
Pluvinage, middle - Le~Sech, bottom - Hylleraas) are shown in the
length, velocity and acceleration (L - solid line, V - dotted line, 
and A - dashed line) gauges.  The
experimental data are from \citet{D96}. }
\end{figure}

In Fig.~\ref{F1} we show the double-to-single photoionization
cross-section ratio for He from threshold to intermediate energies
where it is highly sensitive to electron correlation in the ground
state. We performed calculations with three different ground state
wave functions: Pluvinage, Le~Sech and Hylleraas.  Should the ground
and final states be exact, the calculations in the three gauges of the
electromagnetic operator, the length, velocity and acceleration, would be
identical. Numerical difference between the three gauges is an
indicator of the lack of accuracy in the wave functions. As the final state is
identical in all three calculations, gauge-dependence
indicates the inaccuracy of the ground state. The gauge convergence is
worst for the Pluvinage ground state.  Only the acceleration gauge,
which takes most of its strength at small distances near the nucleus,
gives good results. The two other gauges, the length and velocity,
which are saturated at large and intermediate distances respectively,
are in strong disagreement with each other and the experiment
\cite{D96}. The Le~Sech 
ground state brings significant reduction in the gauge variation,
particularly at the energy of interest to us here of 20~eV above
threshold. The 
Hylleraas ground state yields excellent agreement between the three
gauges and the experiment.

\begin{figure}

\epsffile{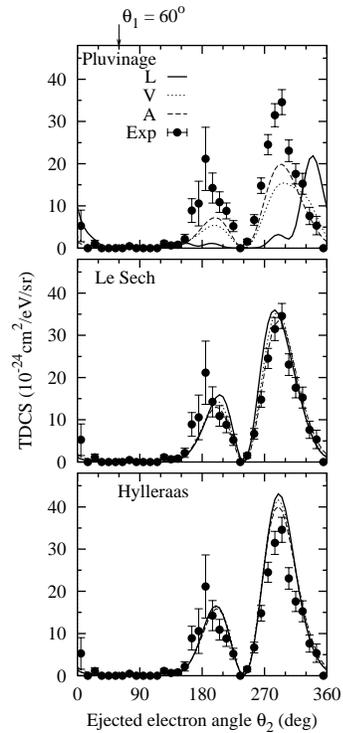}
\caption{
\label{F2}
Triply differential cross section (TDCS) in the polarization plane of
light for ($\gamma$,2e) on He at $E_1=E_2=10$~eV and the Stokes
parameter $S_1=0.98$ As in Fig.~\ref{F1}, the CCC calculations with
different ground states (top - Pluvinage, middle - Le~Sech, bottom -
Hylleraas) are shown in the length, velocity and acceleration (L, V,
and A) gauges. The line styles are as in Fig.~\ref{F1}. The
experimental data are from \citet{BDC98}. }
\end{figure}

To investigate the ground state effects further, we calculate the
angular distribution of the two equal energy photoelectrons
$E_1=E_2=10$~eV in the form of the triply differential cross-section
(TDCS) which is obtained from FDCS \eref{FDCS} in the optical limit
$q\rightarrow0$. This TDCS is the counterpart of the FDCS reported in
the experiments of \citet{ALB99}. In Fig.~\ref{F2} we show the TDCS
for a fixed angle of one of the photoelectrons whereas the second
electron is detected on the full angular range.  We choose
$\theta_1=60^\circ$ where the TDCS is largest in magnitude.  As in the
case of the total DPI cross-section, the Pluvinage ground state gives
the worst results, especially in the length form. Gauge invariance and
agreement with experiment is much better with the Le~Sech and Hylleraas
ground states.

\begin{figure}

\epsffile{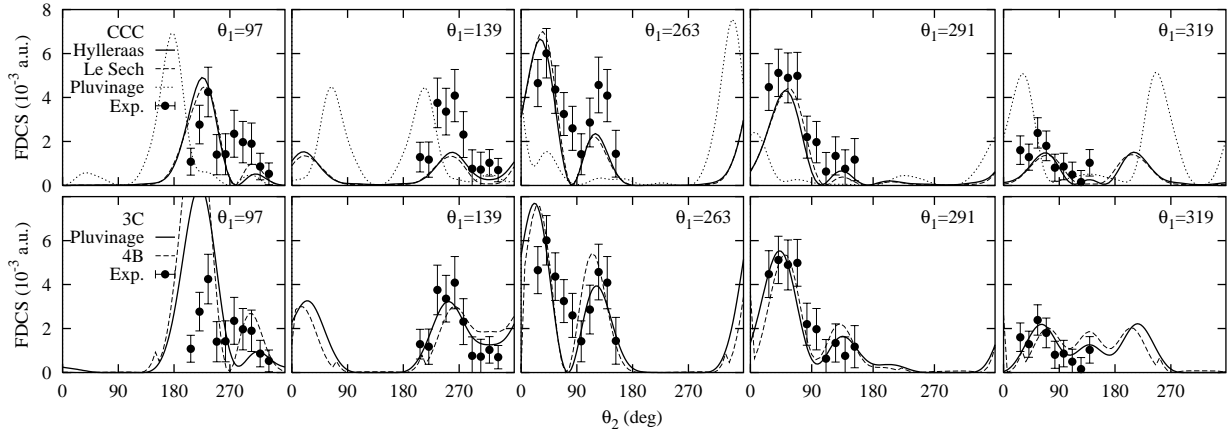}
\caption{
\label{F3}
Fully differential cross section (FDCS) in the scattering plane for
(e,3e) on He at $E_0= 5.6$~keV, $E_1=E_2=10$~eV and $q=0.24$ a.u. for
selected fixed ejected electron angles $\theta_1$. The CCC calculations
with different ground state wave functions are shown on the top row of
panels. The Pluvinage, Le~Sech and Hylleraas ground state results are
shown by the dotted, dashed and solid lines and multiplied by factors
of 6, 1.8 and 2.2, respectively, to match the experiment of
\citet{ALB99}. On the bottom row of panels we show the Pluvinage-3C
calculation of \citet{JM03} and the four-body (4B) Green's function
calculation of \citet{B00}.
}
\end{figure}

Complete failure of the length gauge with the Pluvinage ground state
in DPI has severe implications for (e,3e) calculations which are
also performed in the length gauge of the Born operator \eref{born}.
Results of the CCC calculations of the (e,3e) FDCS is shown on the top
row of panels in Fig.~\ref{F3}. As in Figs.~\ref{F1} and \ref{F2}, we
show calculations with three ground state wave functions: Pluvinage,
Le~Sech and Hylleraas. In addition, the calculations were performed
utilising the first and second Born approximations \cite{K04} and
found to be indistinguishable (similar conclusion was reached in
Ref.~\cite{RWMWM02}). We choose the fixed ejected electron angles at
the values where the four-body Green's function calculations of
\citet{B00} are available. Results of this calculation
along with the 3C-Pluvinage calculation of \citet{JM03} are shown on
the bottom panels.

In the top panel, scaling factors of 2.2 and 1.8 have to be applied to the CCC
calculations with the Hylleraas and Le~Sech ground states, respectively,
to match the experiment. The shape of the FDCS calculated with the
Pluvinage ground state is very different from the experiment and the other
two calculations. A scaling factor of six was applied to this
calculation to put it approximately on the same scale as the experiment.
On the other hand, in the bottom panel,
the two 3C-based calculations of \citet{B00}
and \citet{JM03} agree between themselves and with the absolute
measurements of \citet{ALB99} without any additional scaling. However,
we must recall that simply changing the charge of the projectile
will destroy the good agreement between the two 3C-based calculations.

Thus, we encounter the extraordinary situation where we have confidence in
our results even though they disagree with experiment and two other
theories, all of which agree with each other. We do not agree with the
implicit conclusion of \citet{B00} that the first Born approximation is
insufficient for the problem at hand. While \citet{JM03} also argue
this, they use the Pluvinage ground state which we have shown here not
to work for DPI or (e,3e) when combined with the CCC final
state. However, when this ground state was improved to yield the
Le~Sech state, which still satisfies the Kato's cusp conditions, the
results yielded good agreement with the originally used Hylleraas
ground state. We do not accept that the 3C final state can work better
than the CCC final state for the kinematics considered irrespective of
the initial state. Finally, we have checked the absolute relations between
the  ($\gamma$,2e) and (e,3e) calculations and find them
consistent. Consequently, we stand by the original approach to the
problem \cite{KBDLT99}. Fundamentally, we argue that the physics of
the (e,3e) process of the type considered here is closely related to
the corresponding  ($\gamma$,2e) process. We are hopeful that this
work will stimulate 
further experimental and theoretical study of the subject.
\begin{acknowledgments}

The authors benefited from useful exchange of ideas with Steve Jones,
Don Madison and Claude Dal Cappello.  The Australian Partnership for
Advanced Computing is acknowledged for providing access to the Compaq
AlphaServer SC National Facility.
\end{acknowledgments}


\begin{thebibliography}{20}
\expandafter\ifx\csname natexlab\endcsname\relax\def\natexlab#1{#1}\fi
\expandafter\ifx\csname bibnamefont\endcsname\relax
  \def\bibnamefont#1{#1}\fi
\expandafter\ifx\csname bibfnamefont\endcsname\relax
  \def\bibfnamefont#1{#1}\fi
\expandafter\ifx\csname citenamefont\endcsname\relax
  \def\citenamefont#1{#1}\fi
\expandafter\ifx\csname url\endcsname\relax
  \def\url#1{\texttt{#1}}\fi
\expandafter\ifx\csname urlprefix\endcsname\relax\def\urlprefix{URL }\fi
\providecommand{\bibinfo}[2]{#2}
\providecommand{\eprint}[2][]{\url{#2}}

\bibitem[{\citenamefont{Bray and Stelbovics}(1992)}]{BS92}
\bibinfo{author}{\bibfnamefont{I.}~\bibnamefont{Bray}} \bibnamefont{and}
  \bibinfo{author}{\bibfnamefont{A.~T.} \bibnamefont{Stelbovics}},
  \bibinfo{journal}{Phys.~Rev.~A} \textbf{\bibinfo{volume}{46}},
  \bibinfo{pages}{6995} (\bibinfo{year}{1992}).

\bibitem[{\citenamefont{Madsen and Taulbjerg}(1995)}]{MT95}
\bibinfo{author}{\bibfnamefont{L.~B.} \bibnamefont{Madsen}} \bibnamefont{and}
  \bibinfo{author}{\bibfnamefont{K.}~\bibnamefont{Taulbjerg}},
  \bibinfo{journal}{Phys.~Rev.~A} \textbf{\bibinfo{volume}{52}},
  \bibinfo{pages}{2429} (\bibinfo{year}{1995}).

\bibitem[{\citenamefont{Dewangan}(1997)}]{D97}
\bibinfo{author}{\bibfnamefont{D.~P.} \bibnamefont{Dewangan}},
  \bibinfo{journal}{J.~Phys.~B} \textbf{\bibinfo{volume}{30}},
  \bibinfo{pages}{L467} (\bibinfo{year}{1997}).

\bibitem[{\citenamefont{Yalim et~al.}(1997)\citenamefont{Yalim, Cvejanovic, and
  Crowe}}]{YCC97}
\bibinfo{author}{\bibfnamefont{H.}~\bibnamefont{Yalim}},
  \bibinfo{author}{\bibfnamefont{D.}~\bibnamefont{Cvejanovic}},
  \bibnamefont{and} \bibinfo{author}{\bibfnamefont{A.}~\bibnamefont{Crowe}},
  \bibinfo{journal}{Phys.~Rev.~Lett.} \textbf{\bibinfo{volume}{79}},
  \bibinfo{pages}{2951} (\bibinfo{year}{1997}).

\bibitem[{\citenamefont{O'Neill et~al.}(1998)\citenamefont{O'Neill, {van der
  Burgt}, Dziczek, Bowe, Chwirot, and Slevin}}]{Oetal98}
\bibinfo{author}{\bibfnamefont{R.~W.} \bibnamefont{O'Neill}},
  \bibinfo{author}{\bibfnamefont{P.~J.~M.} \bibnamefont{{van der Burgt}}},
  \bibinfo{author}{\bibfnamefont{D.}~\bibnamefont{Dziczek}},
  \bibinfo{author}{\bibfnamefont{P.}~\bibnamefont{Bowe}},
  \bibinfo{author}{\bibfnamefont{S.}~\bibnamefont{Chwirot}}, \bibnamefont{and}
  \bibinfo{author}{\bibfnamefont{J.~A.} \bibnamefont{Slevin}},
  \bibinfo{journal}{Phys.~Rev.~Lett.} \textbf{\bibinfo{volume}{80}},
  \bibinfo{pages}{1630} (\bibinfo{year}{1998}).

\bibitem[{\citenamefont{Kheifets and Bray}(1998)}]{KB98l}
\bibinfo{author}{\bibfnamefont{A.~S.} \bibnamefont{Kheifets}} \bibnamefont{and}
  \bibinfo{author}{\bibfnamefont{I.}~\bibnamefont{Bray}},
  \bibinfo{journal}{Phys.~Rev.~Lett.} \textbf{\bibinfo{volume}{81}},
  \bibinfo{pages}{4588} (\bibinfo{year}{1998}).

\bibitem[{\citenamefont{Achler et~al.}(2001)\citenamefont{Achler, Mergel,
  Spielberger, D{\"{o}}rner, Azuma, and Schmidt-B{\"o}cking}}]{A01}
\bibinfo{author}{\bibfnamefont{M.}~\bibnamefont{Achler}},
  \bibinfo{author}{\bibfnamefont{V.}~\bibnamefont{Mergel}},
  \bibinfo{author}{\bibfnamefont{L.}~\bibnamefont{Spielberger}},
  \bibinfo{author}{\bibfnamefont{R.}~\bibnamefont{D{\"{o}}rner}},
  \bibinfo{author}{\bibfnamefont{Y.}~\bibnamefont{Azuma}}, \bibnamefont{and}
  \bibinfo{author}{\bibfnamefont{H.}~\bibnamefont{Schmidt-B{\"o}cking}},
  \bibinfo{journal}{J.~Phys.~B} \textbf{\bibinfo{volume}{34}},
  \bibinfo{pages}{965} (\bibinfo{year}{2001}).

\bibitem[{\citenamefont{Kheifets et~al.}(1999)\citenamefont{Kheifets, Bray,
  Duguet, Lahmam-Bennani, and Taouil}}]{KBDLT99}
\bibinfo{author}{\bibfnamefont{A.~S.} \bibnamefont{Kheifets}},
  \bibinfo{author}{\bibfnamefont{I.}~\bibnamefont{Bray}},
  \bibinfo{author}{\bibfnamefont{A.}~\bibnamefont{Duguet}},
  \bibinfo{author}{\bibfnamefont{A.}~\bibnamefont{Lahmam-Bennani}},
  \bibnamefont{and} \bibinfo{author}{\bibfnamefont{I.}~\bibnamefont{Taouil}},
  \bibinfo{journal}{J.~Phys.~B} \textbf{\bibinfo{volume}{32}},
  \bibinfo{pages}{5047} (\bibinfo{year}{1999}).

\bibitem[{\citenamefont{Lahmam-Bennani
  et~al.}(1999)\citenamefont{Lahmam-Bennani, Taouil, Duguet, Lecas, Avaldi, and
  Berakdar}}]{ALB99}
\bibinfo{author}{\bibfnamefont{A.}~\bibnamefont{Lahmam-Bennani}},
  \bibinfo{author}{\bibfnamefont{I.}~\bibnamefont{Taouil}},
  \bibinfo{author}{\bibfnamefont{A.}~\bibnamefont{Duguet}},
  \bibinfo{author}{\bibfnamefont{M.}~\bibnamefont{Lecas}},
  \bibinfo{author}{\bibfnamefont{L.}~\bibnamefont{Avaldi}}, \bibnamefont{and}
  \bibinfo{author}{\bibfnamefont{J.}~\bibnamefont{Berakdar}},
  \bibinfo{journal}{Phys.~Rev.~A} \textbf{\bibinfo{volume}{59}},
  \bibinfo{pages}{3548} (\bibinfo{year}{1999}).

\bibitem[{\citenamefont{Berakdar}(2000)}]{B00}
\bibinfo{author}{\bibfnamefont{J.}~\bibnamefont{Berakdar}},
  \bibinfo{journal}{Phys.~Rev.~Lett.} \textbf{\bibinfo{volume}{85}},
  \bibinfo{pages}{4036} (\bibinfo{year}{2000}).

\bibitem[{\citenamefont{Jones and Madison}(2003)}]{JM03}
\bibinfo{author}{\bibfnamefont{S.}~\bibnamefont{Jones}} \bibnamefont{and}
  \bibinfo{author}{\bibfnamefont{D.~H.} \bibnamefont{Madison}},
  \bibinfo{journal}{Phys.~Rev.~Lett.} \textbf{\bibinfo{volume}{91}}
  (\bibinfo{year}{2003}).

\bibitem[{\citenamefont{Kato}(1957)}]{Kato57}
\bibinfo{author}{\bibfnamefont{T.}~\bibnamefont{Kato}},
  \bibinfo{journal}{Commun. Pure Appl. Math.} \textbf{\bibinfo{volume}{10}},
  \bibinfo{pages}{151} (\bibinfo{year}{1957}).

\bibitem[{\citenamefont{Pluvinage}(1950)}]{Pluvinage50}
\bibinfo{author}{\bibfnamefont{P.}~\bibnamefont{Pluvinage}},
  \bibinfo{journal}{Ann. Phys. (N.Y.)} \textbf{\bibinfo{volume}{5}},
  \bibinfo{pages}{145} (\bibinfo{year}{1950}).

\bibitem[{\citenamefont{Kheifets}(2004)}]{K04}
\bibinfo{author}{\bibfnamefont{A.~S.} \bibnamefont{Kheifets}},
  \bibinfo{journal}{Phys.~Rev.~A} p. \bibinfo{pages}{submitted}
  (\bibinfo{year}{2004}).

\bibitem[{\citenamefont{Moumeni et~al.}(1990)\citenamefont{Moumeni, Dulieu, and
  Sech}}]{MDleS90}
\bibinfo{author}{\bibfnamefont{A.}~\bibnamefont{Moumeni}},
  \bibinfo{author}{\bibfnamefont{O.}~\bibnamefont{Dulieu}}, \bibnamefont{and}
  \bibinfo{author}{\bibfnamefont{C.~L.} \bibnamefont{Sech}},
  \bibinfo{journal}{J.~Phys.~B} \textbf{\bibinfo{volume}{23}},
  \bibinfo{pages}{L739} (\bibinfo{year}{1990}).

\bibitem[{\citenamefont{Siebbeles et~al.}(1993)\citenamefont{Siebbeles,
  Marshall, and Sech}}]{SMleS93}
\bibinfo{author}{\bibfnamefont{L.~D.~A.} \bibnamefont{Siebbeles}},
  \bibinfo{author}{\bibfnamefont{D.~P.} \bibnamefont{Marshall}},
  \bibnamefont{and} \bibinfo{author}{\bibfnamefont{C.~L.} \bibnamefont{Sech}},
  \bibinfo{journal}{J.~Phys.~B} \textbf{\bibinfo{volume}{26}},
  \bibinfo{pages}{L321} (\bibinfo{year}{1993}).

\bibitem[{\citenamefont{Bethe and Salpeter}(1977)}]{BS77}
\bibinfo{author}{\bibfnamefont{H.~A.} \bibnamefont{Bethe}} \bibnamefont{and}
  \bibinfo{author}{\bibfnamefont{E.~E.} \bibnamefont{Salpeter}},
  \emph{\bibinfo{title}{Quantum mechanics of one-and two-electron atoms}}
  (\bibinfo{publisher}{Plenum}, \bibinfo{address}{New York},
  \bibinfo{year}{1977}).

\bibitem[{\citenamefont{Bray}(2002)}]{B02l}
\bibinfo{author}{\bibfnamefont{I.}~\bibnamefont{Bray}},
  \bibinfo{journal}{Phys.~Rev.~Lett.} \textbf{\bibinfo{volume}{89}},
  \bibinfo{pages}{273201} (\bibinfo{year}{2002}).

\bibitem[{\citenamefont{D{\"{o}}rner et~al.}(1996)\citenamefont{D{\"{o}}rner,
  Vogt, Mergel, Khemliche, Kravis, and Cocke}}]{D96}
\bibinfo{author}{\bibfnamefont{R.}~\bibnamefont{D{\"{o}}rner}},
  \bibinfo{author}{\bibfnamefont{T.}~\bibnamefont{Vogt}},
  \bibinfo{author}{\bibfnamefont{V.}~\bibnamefont{Mergel}},
  \bibinfo{author}{\bibfnamefont{H.}~\bibnamefont{Khemliche}},
  \bibinfo{author}{\bibfnamefont{S.}~\bibnamefont{Kravis}}, \bibnamefont{and}
  \bibinfo{author}{\bibfnamefont{C.~L.} \bibnamefont{Cocke}},
  \bibinfo{journal}{Phys.~Rev.~Lett.} \textbf{\bibinfo{volume}{76}},
  \bibinfo{pages}{2654} (\bibinfo{year}{1996}).

\bibitem[{\citenamefont{{Br\"auning} et~al.}(1998)\citenamefont{{Br\"auning},
  {D\"orner}, Cocke, Prior, {Kr\"assig}, Kheifets, Bray, {Br\"auning}-Demian,
  Carnes, Dreuil et~al.}}]{BDC98}
\bibinfo{author}{\bibfnamefont{H.}~\bibnamefont{{Br\"auning}}},
  \bibinfo{author}{\bibfnamefont{R.}~\bibnamefont{{D\"orner}}},
  \bibinfo{author}{\bibfnamefont{C.~L.} \bibnamefont{Cocke}},
  \bibinfo{author}{\bibfnamefont{M.~H.} \bibnamefont{Prior}},
  \bibinfo{author}{\bibfnamefont{B.}~\bibnamefont{{Kr\"assig}}},
  \bibinfo{author}{\bibfnamefont{A.~S.} \bibnamefont{Kheifets}},
  \bibinfo{author}{\bibfnamefont{I.}~\bibnamefont{Bray}},
  \bibinfo{author}{\bibfnamefont{A.}~\bibnamefont{{Br\"auning}-Demian}},
  \bibinfo{author}{\bibfnamefont{K.}~\bibnamefont{Carnes}},
  \bibinfo{author}{\bibfnamefont{S.}~\bibnamefont{Dreuil}},
  \bibnamefont{et~al.}, \bibinfo{journal}{J.~Phys.~B}
  \textbf{\bibinfo{volume}{31}}, \bibinfo{pages}{5149} (\bibinfo{year}{1998}).

\bibitem[{\citenamefont{Rasch et~al.}(2002)\citenamefont{Rasch, Walters,
  Marchalant, Whelan, and Madison}}]{RWMWM02}
\bibinfo{author}{\bibfnamefont{J.}~\bibnamefont{Rasch}},
  \bibinfo{author}{\bibfnamefont{H.~R.~J.} \bibnamefont{Walters}},
  \bibinfo{author}{\bibfnamefont{P.}~\bibnamefont{Marchalant}},
  \bibinfo{author}{\bibfnamefont{C.}~\bibnamefont{Whelan}}, \bibnamefont{and}
  \bibinfo{author}{\bibfnamefont{D.~H.} \bibnamefont{Madison}}, in
  \emph{\bibinfo{booktitle}{Photonic, Electronic, and Atomic Collisions (XXII
  ICPEAC)}}, edited by
  \bibinfo{editor}{\bibfnamefont{J.}~\bibnamefont{Burgdorfer}},
  \bibinfo{editor}{\bibfnamefont{J.}~\bibnamefont{Cohen}},
  \bibinfo{editor}{\bibfnamefont{S.}~\bibnamefont{Datz}}, \bibnamefont{and}
  \bibinfo{editor}{\bibfnamefont{C.}~\bibnamefont{Vane}}
  (\bibinfo{publisher}{Rinton Press}, \bibinfo{address}{Princeton, USA},
  \bibinfo{year}{2002}), pp. \bibinfo{pages}{448--459}.

\end{thebibliography}

\newpage
\widetext

\end{document}